\documentclass[12pt,showpacs]{revtex4} 
\usepackage{amssymb,epsf}
\usepackage{latexsym}
\usepackage{color}
\begin{document}

\title{\textbf{Quantum Corrections for a Braneworld Black Hole}}
\author{Mubasher Jamil}
\email{mjamil@camp.nust.edu.pk} \affiliation{Center for Advanced
Mathematics and Physics, National University of Sciences and
Technology, H-12, Islamabad, Pakistan}

\author{Farhad Darabi}
\email{f.darabi@azaruniv.edu, } \affiliation{Department of Physics,
Azarbaijan University of Tarbiat Moallem, Tabriz, Iran \\
Research Institute for Astronomy and Astrophysics of Maragha (RIAAM), Maragha 55134-441, Iran\\
Corresponding Author}

\date{\today}
\begin{abstract}
\textbf{Abstract:} 
By using the quantum tunneling approach over semiclassical approximations, we study the quantum corrections to the Hawking temperature, entropy and Bekenstein-Hawking entropy-area relation for a black hole lying on a brane.
\end{abstract}
\maketitle
\newpage
\section{Introduction}

There has been widespread interest and activity in the braneworld
gravity in recent years (see \cite{roy1} for a review). The two well-known models of braneworld gravity are those of Randall and Sundrum \cite{ran} and Arkani-Hamed et al \cite{arkani1, arkani2}. 
The braneworld scenario of our universe opens up the fascinating
possibility of the existence of large extra spatial dimensions by
ensuring that the standard model fields are confined to the 3-brane,
whereas gravity could also propagate into the higher dimensional
bulk as well as on the brane and the size of the extra dimensions can be much larger than the Planck length scale. 
The large size of the extra dimension plays the key role in providing a unification scale of the order of a few TeV. The higher dimensional TeV-sized black holes may be then produced at high energy probes. These black holes are microscopic, comparable in the size to the elementary particles. On the other hand, if matter on a 3-brane collapses under gravity a black hole is formed. Then the metric on the braneworld should be close to the Schwarzschild metric at astrophysical scales in order to preserve the observationally tested predictions of general relativity. Braneworld black hole solutions have been studied by a number of authors \cite{roy, Haw, heyd} and the quantum corrections
to some of these solutions have been calculated by different methods. The
possibility that such objects can be produced and detected at the LHC has
been studied in \cite{Casadio1, Casadio2, Casadio3, Casadio4, Casadio5, Casadio6,
Casadio7, Casadio8}. String theory, loop quantum gravity and noncommutative geometry show that in the entropy-area relation the leading order correction should be of log-area type \cite{Str1, Str2, Str3, Str4, Str5, Med, Set1,
Set2, Set3}. On the other hand, generalized uncertainty principle (GUP) and modified dispersion relations (MDRs) provide perturbational framework for such modifications \cite{Noz, Sef, Far, Set4, Set5}.
In the present paper, we use quantum tunneling approach over semiclassical approximations to calculate the quantum corrections to the Hawking temperature and entropy of a Braneworld black hole corresponding to the solution presented in \cite{roy} .

\section{The Model}

The line element of a spherically symmetric black hole of mass $M$
on the brane is \cite{roy,maju}
\begin{equation}\label{1}
ds^2=-\Big[1-\Big( \frac{2M}{M_p^2} \Big)\frac{1}{r}+\Big(
\frac{q}{\tilde {M}_p^2} \Big)\frac{1}{r^2}\Big]dt^2+\Big[1-\Big(
\frac{2M}{M_p^2} \Big)\frac{1}{r}+\Big( \frac{q}{\tilde {M}_p^2}
\Big)\frac{1}{r^2}\Big]^{-1}dr^2+r^2(d\theta^2+\sin^2\theta
d\phi^2),
\end{equation}
Here $M_p$ is the Planck mass while the tilde refers to quantities
on the brane. Also $q=Q\tilde {M}_p^2$ is the dimensionless tidal
charge parameter. Notice that (\ref{1}) represents an exact
localized black hole solution which remarkably has the mathematical
form of the Reissner-Nordstr\"{o}m solution, but without electric
charge being present. Instead the Reissner-Nordstr\"{o}m type
correction can be thought of as a tidal charge which arises from the
projection onto the brane of free gravitational effects in the bulk.

The four dimensional structure of the brane-world black hole depends
on the sign of $q$. For $q\geq0$, there is a direct analogy to the
Reissner-Nordstr\"{o}m solution, with two horizons:
\begin{equation}\label{2}
r_\pm=\frac{M}{M_p^2}\Big[ 1\pm\sqrt{1-q\frac{M_p^4}{M^2\tilde
{M}_p^2}}  \Big].
\end{equation}
To calculate the corrections to the entropy and temperature of
brane-world black hole, we use the Hamilton-Jacobi method to compute
the imaginary part of the action outside the semi-classical
approximation admitting all possible quantum corrections . The
expression of quantum correction of a general function $S(r,t)$
expanded in the series in powers of $\hbar$ is \cite{jhep}
\begin{equation}\label{3}
S(r,t)=S_0(r,t)+\hbar S_1(r,t)+\hbar^2S_2(r,t)+\ldots=S_0(r,t)
+\sum_i\hbar^iS_i(r,t).
\end{equation}
Here $S_0$ is the semiclassical entropy while the remaining terms
are the higher order corrections to $S_0$. On the considerations of
dimensional analysis, the above expression (\ref{3}) takes the form
\cite{sharif}:
\begin{equation}\label{4}
S(r,t)=S_0(r,t)+\sum_i \alpha_i \frac{\hbar^i}{M^{2i}}S_0(r,t)=S_0(r,t)\Big(
1+\sum_i\alpha_i \frac{\hbar^i}{M^{2i}}\Big),
\end{equation}
where $M$ is the mass of the black hole and $\alpha$'s are the
correction coefficients.

The modified form of the temperature of the BH can be written as
\cite{jhep}
\begin{equation}\label{5}
T=T_H\Big( 1+\sum_i\alpha_i \frac{\hbar^i}{M^{2i}} \Big)^{-1},
\end{equation}
where $T_H$ is the standard semiclassical Hawking temperature and
the terms with $\alpha_i$ are corrections due to quantum effects. If
we consider $\alpha_i$ in terms of a single dimensionless
parameter $\beta$ such that $\alpha_i=\beta$ then we have
\begin{equation}\label{5a}
 1+\sum_i\alpha_i \frac{\hbar^i}{M^{2i}}=\Big( 1-\frac{\beta\hbar}{M^2} \Big)^{-1}.
\end{equation}
By using Eq.(\ref{5a}) in Eq.(\ref{5}), we obtain
\begin{equation}\label{5b}
T=T_H\Big( 1-\frac{\beta\hbar}{M^2} \Big),
\end{equation}
where the Hawking temperature of the braneworld black hole is given by
\begin{equation}\label{6}
T_H=\frac{\hbar}{4\pi}\frac{\partial g_{00}}{\partial r}|_{r=r_+}=
\frac{\hbar}{2\pi}\Big( \frac{M\tilde {M}_p^2r_+-qM_p^2}{\tilde
{M}_p^2M_p^2r_+^3} \Big).
\end{equation}
Therefore, the semiclassical entropy of the braneworld black hole is obtained
as follows
\begin{equation}\label{7}
S_0=\int\frac{1}{T_H}dM=\frac{2\pi}{\hbar}\int \Big( \frac{\tilde
{M}_p^2M_p^2r_+^3}{M\tilde {M}_p^2r_+-qM_p^2} \Big)dM.
\end{equation}
The mass of the braneworld black hole is given by
\begin{equation}\label{8}
M=\frac{M_p^2r_+}{2}\Big( 1+\frac{q}{\tilde {M}_p^2}\frac{1}{r_+^2}
\Big),
\end{equation}
while the corresponding differential of mass is easily obtained
\begin{equation}\label{9}
dM=\Big(\frac{M_p^2\tilde {M}_p^2r_+^2-qM_p^2}{2\tilde
{M}_p^2r_+^2}\Big)dr_+.
\end{equation}
Making use of (\ref{8}) and (\ref{9}) in (\ref{7}) yields
\begin{equation}\label{10}
S_0=\frac{2\pi}{\hbar}M_p^2\int r_+dr_+=\frac{\pi}{\hbar}M_p^2r_+^2.
\end{equation}
Assuming $A$ is the area of the event horizon, it is then calculated as follows
\begin{equation}\label{10a}
A=\int\sqrt{g_{22}g_{33}}dx^2dx^3=4\pi r_+^2.
\end{equation}
Thus, Eq.(\ref{10}) can be written as
\begin{equation}\label{11}
S_0=\frac{A}{4\hbar}M_p^2,
\end{equation}
and we obtain the Hawking entropy-area relation for a braneworld
black hole.

\section{Hawking Temperature Corrections}

In this section, we find the correction to the Hawking temperature
as a result of quantum effects for the braneworld black hole. The
expression for the semiclassical Hawking temperature, (\ref{6})
turns out to be
\begin{equation}\label{12}
T_H= \frac{\hbar}{4\pi}\Big( \frac{\tilde {M}_p^2r_+^2-q}{\tilde
{M}_p^2r_+^3} \Big).
\end{equation}
On the other hand, equations (\ref{5b}) and (\ref{8}) result in
\begin{equation}\label{13}
T=T_H\Big[
1-\beta\hbar\frac{4\tilde{M}_p^4r_+^2}{(M_p^2\tilde{M}_p^2r_+^2+qM_p^2)^2}
\Big].
\end{equation}
Now using (\ref{12}) in (\ref{13}), the corrected Hawking temperature is given by
\begin{equation}\label{14}
T=\frac{\hbar}{4\pi}\Big( \frac{\tilde {M}_p^2r_+^2-q}{\tilde
{M}_p^2r_+^3} \Big)\Big[
1-\beta\hbar\frac{4\tilde{M}_p^4r_+^2}{(M_p^2\tilde{M}_p^2r_+^2+qM_p^2)^2}
\Big].
\end{equation}

\section{Entropy Corrections}

Here we calculate the quantum corrections to the entropy of the
Braneworld black hole. In terms of the horizon radius, the corrected
form of the entropy (\ref{4}) is given by
\begin{equation}\label{15}
S(r,t)=S_0\Big( 1+\sum_i
\frac{\alpha_i\hbar^i(2\tilde{M}_p^2r_+)^{2i}}{(M_p^2\tilde{M}_p^2r_+^2+qM_p^2)^{2i}}\Big),
\end{equation}
and similarly the corrected form of the Hawking temperature can be
written as
\begin{equation}\label{16}
T=T_H\Big( 1+\sum_i
\frac{\alpha_i\hbar^i(2\tilde{M}_p^2r_+)^{2i}}{(M_p^2\tilde{M}_p^2r_+^2+qM_p^2)^{2i}}\Big)^{-1}.
\end{equation}
In the first law of thermodynamics $dM=TdS$, we replace the
temperature $T$ by the corrected form of the temperature. Thus the
entropy with the correction terms is given by
\begin{equation}\label{17}
S(M)=\int \frac{1}{T_H}\Big( 1+\sum_i
\frac{\alpha_i\hbar^i(2\tilde{M}_p^2r_+)^{2i}}
{(M_p^2\tilde{M}_p^2r_+^2+qM_p^2)^{2i}}\Big)dM,
\end{equation}
which can be written in expanded form as
\begin{eqnarray}\label{18}
S(M)&=&\int \frac{1}{T_H}dM+\int
\frac{\alpha_1\hbar(2\tilde{M}_p^2r_+)^{2}}{T_H(M_p^2\tilde{M}_p^2r_+^2+qM_p^2)^{2}}
dM+\int
\frac{\alpha_2\hbar^2(2\tilde{M}_p^2r_+)^{4}}{T_H(M_p^2\tilde{M}_p^2r_+^2+qM_p^2)^{4}}
dM+\ldots\nonumber\\
&=& I_1+I_2+I_3+\ldots,
\end{eqnarray}
where the first integral $I_1$ has been evaluated in (\ref{7}) and
$I_2$, $I_3,\ldots$ are quantum corrections. Thus,
\begin{equation}\label{19}
I_2=\frac{2^3\pi\alpha_1\tilde{M}_p^4}{M_p^2}
\int\frac{r_+^3}{(\tilde{M}_p^2r_+^2+q)^{2}}dr_+,
\end{equation}
and
\begin{equation}\label{20}
I_3=\frac{2^5\pi\alpha_2\hbar\tilde{M}_p^8}{M_p^6}
\int\frac{r_+^5}{(\tilde{M}_p^2r_+^2+q)^{4}}dr_+,
\end{equation}
where use has been made of (\ref{9}). 
In general, we can write
\begin{equation}\label{21}
I_k=\frac{2^{2k-1}\pi\alpha_{k-1}\hbar^{k-2}
\tilde{M}_p^{4k-6}}{M_p^{4k-2}}\int\frac{r_+^{2k-1}}{(\tilde{M}_p^2r_+^2+q)
^{2k-2}}dr_+,\
\ k>3.
\end{equation}
Therefore the entropy with quantum corrections is given by
\begin{eqnarray}\label{22}
S(M)&=&\frac{2\pi}{\hbar}M_p^2\int r_+dr_+
+\frac{2^3\pi\alpha_1\tilde{M}_p^4}{M_p^2}
\int\frac{r_+^3}{(\tilde{M}_p^2r_+^2+q)^{2}}dr_+\nonumber\\&&+\sum_{k>2}\frac{2^{2k-1}\pi\alpha_{k-1}\hbar^{k-2}
\tilde{M}_p^{4k-6}}{M_p^{4k-2}}\int\frac{r_+^{2k-1}}{(\tilde{M}_p^2r_+^2+q)
^{2k-2}}dr_+.
\end{eqnarray}
After the evaluation of the integrals, (\ref{22}) takes the form
\begin{eqnarray}\label{23}
S(M)&=&\frac{\pi}{\hbar}M_p^2r_+^2+\frac{2^2\pi\alpha_1}{M_p^2}\Big(
\ln(q+r_+^2\tilde{M}_p^2)+\frac{q}{q+r_+^2\tilde{M}_p^2}\Big)\nonumber\\&&
+\sum_{k>2}\frac{2^{2k-2}\pi\alpha_{k-1}\hbar^{k-2}
\tilde{M}_p^{4k-6}}{kM_p^{4k-2}(2+3k+k^2)}\Big[
\frac{r_+^{2k}}{q^{2k}}\Big\{(2+3k+k^2)q^2{}_2F_{1}\Big(
k,2k,1+k,-\frac{r_+^2\tilde{M}_p^2}{q} \Big)\nonumber\\&&
+kr_+^2\tilde{M}_p^2\Big[ 2(2+k)q{}_2F_{1}\Big(
1+k,2k,2+k,-\frac{r_+^2\tilde{M}_p^2}{q} \Big)\nonumber\\&&
+(1+k)r_+^2\tilde{M}_p^2{}_2F_{1}\Big(
2+k,2k,3+k,-\frac{r_+^2\tilde{M}_p^2}{q} \Big) \Big] \Big\} \Big].
\end{eqnarray}
The entropy in (\ref{23}) in terms of area $A$ is given as follows:
\begin{eqnarray}\label{24}
S(M)&=&\frac{A}{4\hbar}M_p^2+\frac{2^2\pi\alpha_1}{M_p^2}\Big(
\ln(q+\frac{A}{4\pi}\tilde{M}_p^2)+\frac{q}{q+\frac{A}{4\pi}\tilde{M}_p^2}\Big)\nonumber\\&&
+\sum_{k>2}\frac{2^{2k-2}\pi\alpha_{k-1}\hbar^{k-2}
\tilde{M}_p^{4k-6}}{kM_p^{4k-2}(2+3k+k^2)}\Big[
\frac{A^k}{\pi^k(2q)^{2k}}\Big\{(2+3k+k^2)q^2{}_2F_{1}\Big(
k,2k,1+k,-\frac{A\tilde{M}_p^2}{4\pi q} \Big)\nonumber\\&&
+k\frac{A}{4\pi}\tilde{M}_p^2\Big[ 2(2+k)q{}_2F_{1}\Big(
1+k,2k,2+k,-\frac{A\tilde{M}_p^2}{4\pi q} \Big)\nonumber\\&&
+(1+k)\frac{A}{4\pi}\tilde{M}_p^2{}_2F_{1}\Big(
2+k,2k,3+k,-\frac{A\tilde{M}_p^2}{4\pi q} \Big) \Big] \Big\} \Big].
\end{eqnarray}
Here ${}_2F_{1}$ is the hypergeometric function. Clearly the first
term in the above expansion (\ref{24}) is the usual Hawking
entropy-area relation. However the first correction term is
logarithmic which incorporates the contribution of the tidal charge.
The higher order corrections follow a combination of hypergeometric
functions. We emphasize that at this stage the parameter $q$ cannot be taken zero since it will make the correction terms divergent.
\newpage
\section{Conclusion}

There are several approaches such as string theory,
loop quantum gravity, noncommutative geometry, modified dispersion relations and generalized uncertainty principle to find quantum gravitational corrections of black hole entropy-area relation. In this paper, using the quantum tunneling approach over semiclassical approximations, we have studied the quantum corrections to the thermodynamical quantities like Hawking temperature, entropy and Bekenstein-Hawking
entropy-area relation of a Braneworld black hole. It is shown that similar
to the other approaches a logarithmic correction, among other corrections,
appears which incorporates the contribution of the dimensionless tidal charge
parameter on the brane.

\section*{Acknowledgment}
This work has been supported by ``Research Institute for
Astronomy and Astrophysics of Maragha (RIAAM)'', Iran.

\end{document}